\documentstyle[sprocl,epsf,epsfig]{article}

\def\be{\begin{equation}}
\def\ee{\end{equation}}
\def\bea{\begin{eqnarray}}
\def\eea{\end{eqnarray}}

\begin{document}

\title{HOW GOOD ARE PRESENT ANALYTICAL QCD-PREDICTIONS 
ON FLUCTUATIONS IN ANGULAR INTERVALS ?}

\author{B.BUSCHBECK and \underline{F.MANDL}}

\address{Institute for High Energy Physics of the OAW, Nikolsdorferstrasze 18, 
\\A-1050 Wien, Austria\\E-mail:mandl@hephy.oeaw.ac.at}

\maketitle\abstracts{ 
Results on two--particle angular correlations in jet cones 
and on multiplicity fluctuations in one- and two- dimensional
angular intervals, delivered by three experiments (DELPHI, L3 and ZEUS, at
$\sqrt{s}$ from few to 183 GeV) are compared to present existing 
analytical QCD calculations,
using the LPHD hypothesis. 
Two different types of functions have been tested. While the differentially
normalized correlation functions show substantial deviations from the 
predictions, a globally normalized correlation function agrees 
surprisingly well.
The role of the QCD parameters $\alpha_s$, $\Lambda$ and $n_f$ is discussed.
The necessity to include full energy-momentum conservation into the
analytical calculations is stressed.
}

%
\makeatletter
\newcount\@tempcntc
\def\@citex[#1]#2{\if@filesw\immediate\write\@auxout{\string\citation{#2}}\fi
  \@tempcnta\z@\@tempcntb\m@ne\def\@citea{}\@cite{\@for\@citeb:=#2\do
    {\@ifundefined
       {b@\@citeb}{\@citeo\@tempcntb\m@ne\@citea\def\@citea{,}{\bf ?}\@warning
       {Citation `\@citeb' on page \thepage \space undefined}}%
    {\setbox\z@\hbox{\global\@tempcntc0\csname b@\@citeb\endcsname\relax}%
     \ifnum\@tempcntc=\z@ \@citeo\@tempcntb\m@ne
       \@citea\def\@citea{,}\hbox{\csname b@\@citeb\endcsname}%
     \else
      \advance\@tempcntb\@ne
      \ifnum\@tempcntb=\@tempcntc
      \else\advance\@tempcntb\m@ne\@citeo
      \@tempcnta\@tempcntc\@tempcntb\@tempcntc\fi\fi}}\@citeo}{#1}}
\def\@citeo{\ifnum\@tempcnta>\@tempcntb\else\@citea\def\@citea{,}%
  \ifnum\@tempcnta=\@tempcntb\the\@tempcnta\else
   {\advance\@tempcnta\@ne\ifnum\@tempcnta=\@tempcntb \else \def\@citea{--}\fi
    \advance\@tempcnta\m@ne\the\@tempcnta\@citea\the\@tempcntb}\fi\fi}
 
\makeatother
\newcommand{\DpName}[2]{\hbox{#1$^{\ref{#2}}$},\hfill}
\newcommand{\DpNameTwo}[3]{\hbox{#1$^{\ref{#2},\ref{#3}}$},\hfill}
\newcommand{\DpNameThree}[4]{\hbox{#1$^{\ref{#2},\ref{#3},\ref{#4}}$},\hfill}
\newskip\Bigfill \Bigfill = 0pt plus 1000fill
\newcommand{\DpNameLast}[2]{\hbox{#1$^{\ref{#2}}$}\hspace{\Bigfill}}
%
%
\newcommand{\gsim}{\;\raisebox{-0.9ex}{$\textstyle\stackrel{\textstyle>}{\sim}$}\;}
\newcommand{\lsim}{\;\raisebox{-0.9ex}{$\textstyle\stackrel{\textstyle<}{\sim}$}\;}

\section{Introduction} Phenomenological models which try to describe 
multihadron production in $e^+e^-$ reactions have to cut off the parton
cascade at some scale $Q_0 \geq 1~GeV$. The following non-perturbative 
hadronisation
phase is "handled" with many parameters such that the connection to
QCD is "diluted".
It has been suggested to extend instead the parton evolution 
calculated using perturbative QCD down to a 
lower mass scale.
The multihadron final states 
are then compared directly with the multiparton final states
\cite{parton} by using 
the concept of Local Parton Hadron
Duality (LPHD)
\cite{lhd1}. The obvious difficulty is the overlap of perturbative and 
non-perturbative QCD. It is unclear what one could expect when entering
a region in which nobody knows what to apply. Therefore an experimental
input is vital and the only way to learn what on really can apply.
The main
theoretical effort for evaluating multiparton correlations in the framework of QCD has
been based on the Double Log Approximation (DLA)
\cite{parton,dla}. 
Comparisons with experimental data have to cope with substantial 
simplifications in the
calculations of the perturbative part which are justified only at asymptotic 
energies, as well as with the 
question of how far the LPHD hypothesis is valid. 

Detailed prescriptions for multiparton angular correlations in cones using
DLA have been proposed by three groups \cite{ochs,pesch,dremin}, which
calculated parton correlations produced in gluon
cascades radiated off the initial parton. 
The analytical predictions considered thereby
involve only one adjustable parameter, namely the QCD scale $\Lambda$.
The probability for gluon
bremsstrahlung in DLA reads as follows: 
\begin{equation} 
M(k)d^3k = c_a \gamma_0^2 \frac{dk}{k}
\frac{d\Theta_{pk}}{\Theta_{pk}}
\frac{d\Phi_{pk}}{2\pi}   \hspace{1.2cm}
\gamma_0^2 = 6\alpha_S/\pi
\end{equation}
where $c_g=1$ and $c_q=\frac{4}{9}$, $p$ and $k$ are the 3-momenta 
of the parent 
parton and the radiated gluon, $\Theta_{pk}$ the angle of emission of the 
gluon, $\Phi_{pk}$ the azimuthal angle of the gluon around $\vec{p}$ and
$\gamma_0$ is dependent on the $p_t$ of the gluon.
The inclusive n-particle densities $\rho_n(k_1, k_2,.....k_n)$ ($k_i$ is 
the 3-momentum of the i-th particle) are obtained by applying the 
generating functional technique
\cite{x1} which has been developed for QCD jets first by \cite{x2,dla}.
The calculations have been carried out in DLA where the integrals involved are 
performed only in phase space regions with dominant contributions given by 
the singularities of (1). Among the simplifications that had to be used are
neglecting energy-momentum conservation and $q\bar{q}$ production, and  
assuming well developed cascades at very high energies. Angular 
ordering, however, has been taken into account.

The predictions of \cite{ochs,pesch,dremin} and the present study use 
the lowest order 
QCD relations 
between the coupling 
$\alpha_s$ and the QCD scale $\Lambda$.
\begin{equation}
\alpha_s = \frac{\pi\beta^2}{6}\frac{1}{\ln(\frac{Q}{\Lambda})} \hspace{1.2cm}
\beta^2 = 12(\frac{11}{3}n_c - \frac{2}{3}n_f)^{-1}
\end{equation}

The aim of this study is to compare experimental measurements of angular
correlations in cones around the jet axis as well as multiplicity fluctuations
in one- and two- dimensional angular intervals with the
available theoretical QCD predictions thus helping the theorists to assess
how well their analytical calculation is and what improvements could
be done.
First experimental measurements \cite{wien2,wien3} revealed 
substantial
deviations, but also encouraging agreements.

This study uses results of the LEP experiments DELPHI \cite{de1,de2} and 
L3 \cite{l3} and
the HERA experiment ZEUS \cite{ze}.
Section 2 contains information about the experimental data used for the
comparison.
In sections 3 and 4 the theoretical framework is sketched and the
analytical calculations are compared with the experimental measurement. 
Section 5 contains the final 
discussion and the summary.

\section{The data samples}
All three experiments considered use specific cuts for hadronic
events and track quality  
(DELPHI \cite{delphi,x}, L3 \cite{l3}, ZEUS \cite{zeus})
Various corrections were applied using events 
generated from Monte Carlo simulations \cite{de1,de2,l3,ze}.  
In the following only corrected data 
will be shown.

\underline{DELPHI}:
The analysis uses about 600k $e^+e^-$ interactions (after cuts) 
at $\sqrt{s} = 91$~GeV .
A sample of about 1200 high energy events at $\sqrt{s} = 183~GeV$ collected 
in 1997
is used to investigate the energy dependence. 

\underline{L3}:
About 1 million $e^+e^-$ interactions 
at $\sqrt{s} = 91$~GeV  are used.

The event axis is the sphericity axis both in DELPHI and L3.

\underline{ZEUS}:
The data of this experiment ($ep$ interactions at HERA) are 4 samples
with different ranges of $Q^2$: 85k events ($10 \leq Q^2 \leq 20$~GeV), 
28k events
($Q^2 \geq 100$~GeV), 0.9k events ($Q^2 \geq 1000$~GeV) and 0.3k events with
($Q^2 \geq 2000$~GeV). The last sample corresponds to $\sqrt{s} \approx 62~GeV$,
which comes near to LEP1 energies. The event axis is defined using the initial
quark momentum of the $\gamma^* \mathrm{q}$ collision in the quark-parton
model.

\section{Multiplicity fluctuations in 1- and 2- dimensional ring regions
around jet cones}
The theoretical calculations treat correlations between partons 
emitted within an angular window 
defined by two angles $\vartheta$ and $\Theta$. 
The parton and particle density  
correlations (fluctuations) in this window are described by 
normalized factorial moments of order $n$:
\begin{equation}
F^{(n)}(\Theta ,\vartheta )  =  \frac{\int 
\prod_{k=1}^{n} d\Omega_k \rho^{(n)} (\Omega_1 , \ldots , 
\Omega_n )} {\int
\prod_{k=1}^{n} d\Omega_k 
\rho^{(1)} (\Omega_k )} 
\end{equation}
The integrals extend over the window chosen. 
The angular windows considered here are either rings around the 
jet axis with mean 
opening angle $\Theta$ and half width $\vartheta$ in the 
case of 1 dimension ($D=1$), or cones with half opening angle $\vartheta$ around
a direction ($\Theta,\Phi$) with respect to the jet axis in the case of
2 dimensions ($D=2$). At sufficiently large jet energies, the parton flow in 
these angular windows is dominated by parton avalaches caused
by gluon bremsstrahlung off the initial quark.

Ref.~\cite{ochs} derived their predictions explicitly for cumulant moments
$C^{(n)}$ \cite{cumu},
whereas~\cite{pesch} and~\cite{dremin} obtained similar expressions for the
factorial moments $F^{(n)}$.  

For the normalized cumulant moments
$C^{(n)}$~\cite{ochs} and the factorial moments
$F^{(n)}$~\cite{pesch,dremin},
the following prediction  has been made:
\begin{equation} C^{(n)}(\Theta,\vartheta )~{\rm{or}}~F^{(n)} (\Theta ,\vartheta ) \sim 
\left(\frac{\Theta}{\vartheta}\right)^{\phi_n}
\end{equation}
All 3 references~\cite{ochs,pesch,dremin} give in the high energy limit 
and for large $\vartheta < \Theta$ the same linear
approximation for the exponents $\phi_n$:
\begin{equation}
\phi_n \approx (n-1)D - \left(n - \frac{1}{n}\right) \gamma_0 
\end{equation}
For fixed $\alpha_s$ (along the parton shower)  eq.~(5) is asymptotically 
valid for all
angles $\vartheta$. 

When the running of $\alpha_S$ with $\vartheta$ in the parton cascade is taken into
account, ref.~\cite{ochs} obtained

\begin{equation}
\phi_n = (n-1)D - 2\gamma_0 (n-\omega (\epsilon , n))/\epsilon
\end{equation}

\noindent where $D=1$ for ring regions and $D=2$ for cones,

\begin{equation} 
\omega(\epsilon ,n) = n\sqrt{1-\epsilon}(1-\frac{1}{2n^2} \ln(1-\epsilon))
\end{equation}     
and
\begin{equation}
\epsilon = \frac{\ln(\Theta/\vartheta)}{\ln(P\Theta/\Lambda)}
\end{equation}
where $P \approx 
\frac{\sqrt{s}}{2}$ is the momentum of the initial parton.

The dependence on the QCD parameters $\alpha_s$ or $\Lambda$ enters in the
above equations via $\gamma_0$ and $\epsilon$ that are determined by the scale
$Q \approx P\Theta$.
In the present study it is, for $\Theta=25^o$,  
about 20 GeV for $\sqrt{s}$=91.1~GeV and
about 38~GeV for $\sqrt{s} \approx 183~GeV$.

The corresponding predictions of refs.~\cite{pesch} (eq.~9) and~\cite{dremin}
(eq.~10) are analytically different, but numerically similar:

\begin{equation}
\phi_n = (n-1)D - \frac{2\gamma_0}{\epsilon} \cdot \frac{n^2 -1}{n} \left(1 -
\sqrt{1-\epsilon}\right)
\end{equation}

\begin{equation}
\phi_n = (n-1)D - \frac{n^2 - 1}{n} \gamma_0 \left( 1 + \frac{n^2 +1} {4 n^2}
\epsilon\right)
\end{equation}

These relations depend also on the number of flavours ($n_f$). Since
equ.~2 emerges only from "one loop" calculations, the parameter $\Lambda$
is not the universal $\Lambda_{\overline{MS}}$, but only
an effective parameter $\Lambda_{eff}$.
But also in this approximation $\alpha_s$ is running having a
scale dependence $1/\ln(Q^2/\Lambda^2)$.
The running of $\alpha_s$ during the process of jet cascading is 
implicitly taken into account in (6), (9) and (10) by the dependence of
$\phi_n$ on $\epsilon$ (or $\vartheta$). In theory this causes a bending 
of $F^{(n)}$ 
when approaching smaller values of $\vartheta$ (larger $\epsilon$).

\newpage
\begin{figure}
\vspace{-1.5cm}
\mbox{\epsfig{file=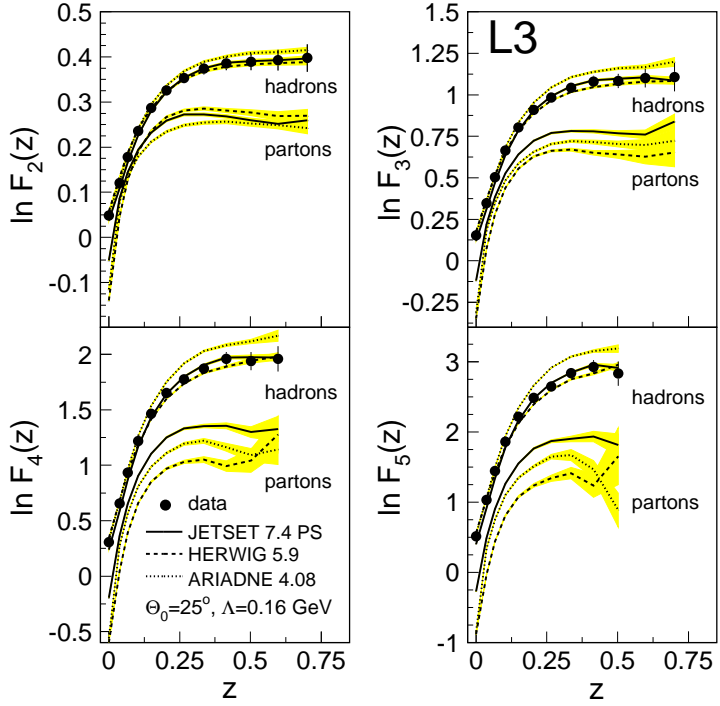,width=10cm}}
\caption[pict1]{Factorial moments in 1-dim rings are 
compared with three
Monte Carlo models, both on hadronic and partonic level.
Please note that the variable 
$\epsilon$ (equ. 8) is here denoted by z.}
\end{figure}  

In fig.~1 (L3) the factorial moments of orders 2,3,4 and 5 are  
compared to various Monte Carlo simulations at hadronic and partonic level.
{\it For all orders n the hadronic levels agree very well with the data 
(this is generally the case), 
while the partonic level disagrees with that of the hadronic level},
so that  LPHD seems to be not valid in this case. 
This is somewhat contradicted by ZEUS \cite{ze}, which claim an 
approximate validity
for LPHD for $Q^2 \geq 100~GeV^2$. 
When normalizing these moments by the
respective value at $\epsilon = 0$ {\it the agreement between parton and
hadron levels is improved \cite{l3}.} 
It is pointed to the fact that the shape of the correlation functions on
the partonic level depends very much on the cut-off parameter $Q_0$ 
\cite{wien3} (in fig.~1 $Q_0$=1~GeV, a rather high value compared to the
original understanding of LPHD \cite{lhd1}).

Fig.~2 (DELPHI) shows the normalized factorial moments of orders 2, 3, 4 
and 5  
together with the predictions of refs.~\cite{ochs,pesch,dremin}, 
in one- and two- dimensional angular intervals (i.e. rings and side cones) for
$\Lambda$= 0.15~GeV  and $n_f$=3. 
{\it They  
are not described well by the theoretical predictions~\cite{ochs,pesch,dremin}}

\begin{figure}
\vspace{-1.5cm}
\mbox{\epsfig{file=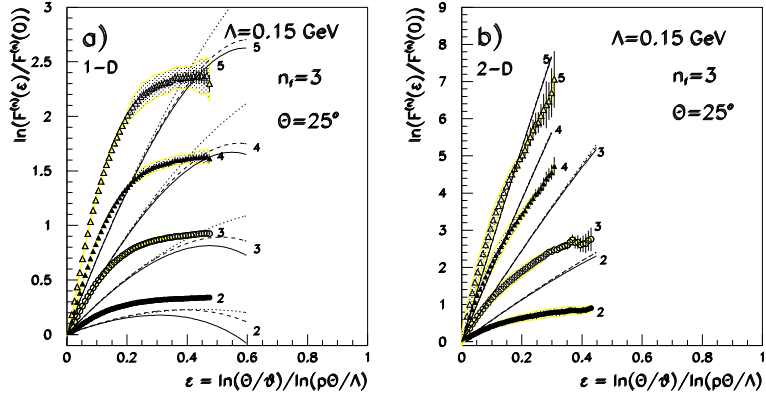,width=11cm}}
\vspace{-1cm}
\caption[pichalf]{
Factorial moments in {\bf a)} 1-dim rings and {\bf b)} 2-dim side cones are
compared with analytical calculations of
refs~\cite{ochs,pesch,dremin}, eqs. (6) (solid lines), (9) (dashed lines),
and (10) (dotted lines), for a cone opening angle of $\Theta = 25^o$. 
As a consistency test, 1-dim and 2-dim factorial moments are
compared with same values of QCD parameters: note the different vertical 
scales. The values for 
orders 2 to 5 are indicated in all figures.
The statistical errors are shown by the error bars,
the shaded regions indicate the systematic errors.
}
\end{figure}

For the 1-dim case (fig.~1a) the predictions lie below the data for not too 
large $\epsilon$, 
differing also in shape. 
Choosing $n_f = 5$  the discrepances will increase \cite{de1}, {\it choosing
smaller values of $\Lambda$ (e.g. =~0.04~GeV) will  
reduce the discrepancies} for small $\epsilon$ \cite{de1,l3}, especially 
for lower orders n. 
The angular correlations in 2 dimensions (fig.~2b) 
{\it show a different (disagreeing) 
behaviour} for the lower order moments $n < 4$, where the predictions lie
above the data. 
The higher moments $F^{(4)}$ and $F^{(5)}$ 
have similar features in the 1-dim and 2-dim case. In both cases the data 
lie above the predictions at small $\epsilon$ and bend below them at larger
$\epsilon$.

Similar conclusions for $\sqrt{s}$=91~GeV, for the 1-dim ring regions, 
are drawn by the 
L3 collaboration.

{\it It is not possible to find \underline{one} set of QCD parameters 
$\Lambda$ and $n_f$
which simultaneously minimize the discrepancies between data and predictions 
for moments
of all orders 2,3,4 and 5 in both the 1-dim. and 2-dim. cases}.  

\begin{figure}
\vspace{-2.5cm}
\begin{minipage}[t]{5.5cm}
\mbox{\epsfig{file=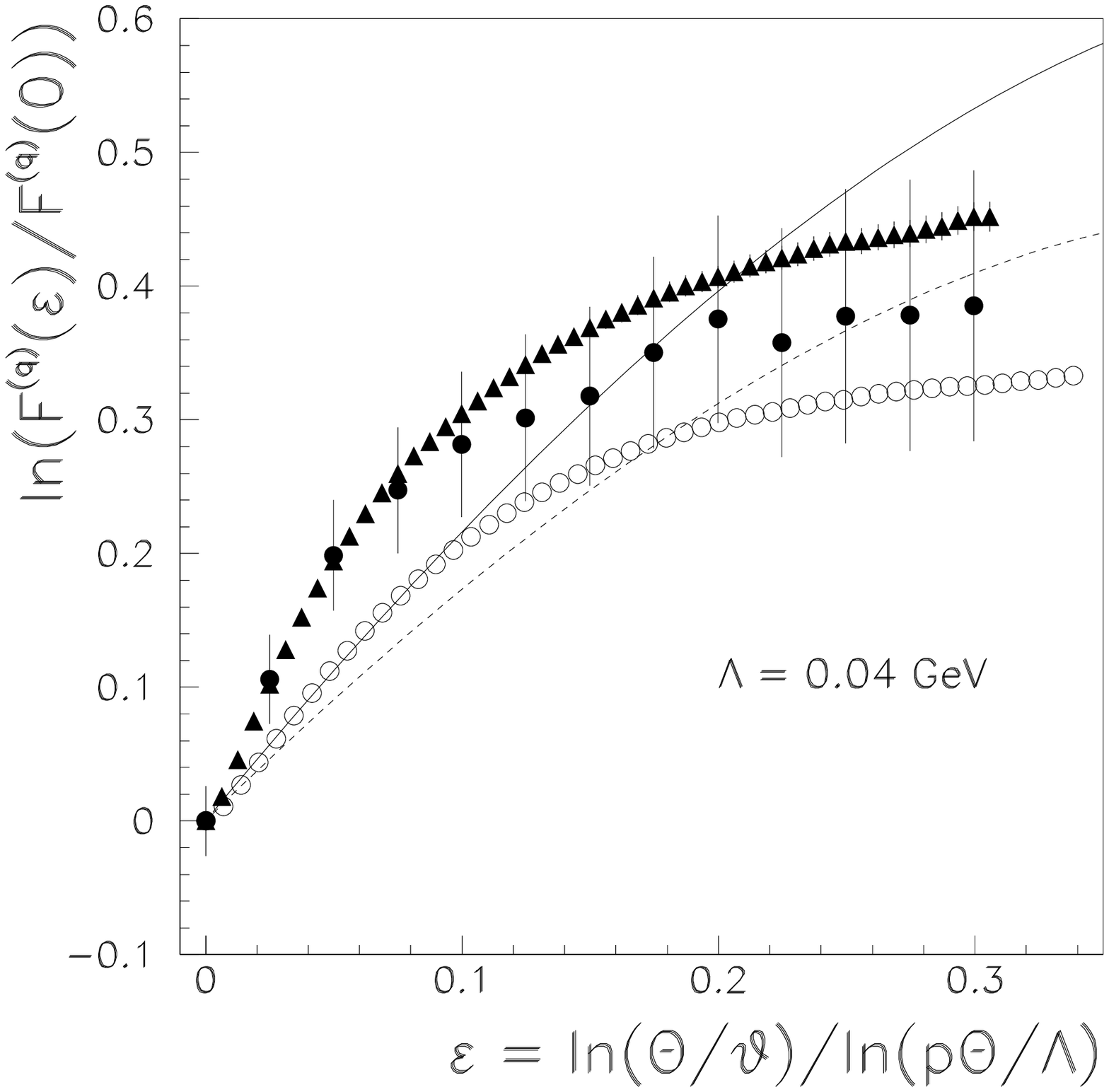,width=5.5cm}}
\vspace{-0.5cm}
\caption[]{ 
\footnotesize 
The energy dependence of the normalized factorial moment of order 2 
$\sqrt{s}$=91.1~GeV: data (open circles), in full agreement with
JETSET, and prediction \cite{ochs} 
dashed line and
$\sqrt{s}$=175~GeV: data (full circles) and prediction \cite{ochs} 
solid line,  
for $n_f=3$ and $\Lambda$=0.04~GeV.
The full triangles denote the high energy JETSET simulation. 
}
\end{minipage} \hfill
\begin{minipage}[t]{5.5cm}
\mbox{\epsfig{file=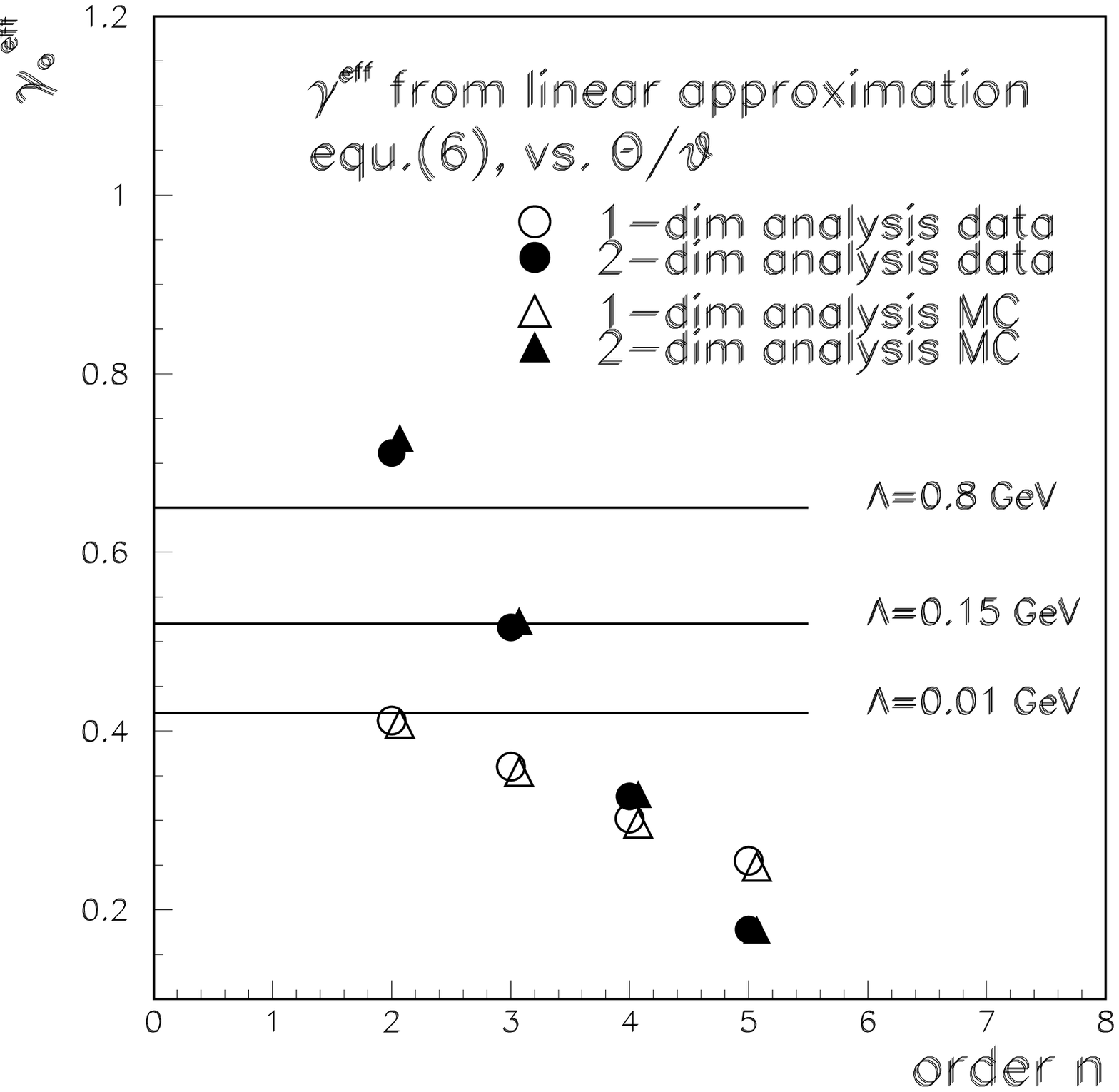,width=5.5cm}}
\vspace{-0.5cm}
\caption[]{
\footnotesize Values of $\gamma^{eff}_0$ obtained from fitting the 
linear approximation 
eq.~(5)  vs. $\frac{\Theta}{\vartheta}$ 
the 1-dim case (open circles) and the 2-dim case (full circles), for
orders $n=2,3,4,5$. The measured values of $\gamma^{eff}_{0}$ 
are also compared 
with those obtained from JETSET at generator level: open triangles 
(1-dim) and full triangles (2-dim).}
\end{minipage}
\end{figure}

Fig.~3 (DELPHI) shows a comparison with high energy data at $\sqrt{s}$=183~GeV 
(with a mean energy of $\sqrt{s}$=175~GeV and the
corresponding predictions according to equ. (6), where the energy dependence
enters via the parameter $\gamma_0$. 
For larger values of $\epsilon$ there is better agreement, 
the statistical
errors of the high energy data, however, are substantial. 
The relative increase of the
predicted moments agrees qualitatively with that of the JETSET model. 


The QCD parameter $\gamma_0$ is discussed in 
fig.~4 (DELPHI), where
the numerical values of $\gamma_0^{eff}$ derived from the measured slopes
$\phi_n$ (by fitting equ. (5) for $\epsilon \leq 0.1$) are given 
for the orders $n=2,3,4,5$.
From the present theoretical understanding,
$\gamma_0$ is expected to be independent of $n$.  
This is indicated by horizontal lines in fig.~4.
It has to be pointed out that the average 
measured values of $\gamma_0^{eff}$ are not too far from the expectation.
{\it The $n$-dependence observed, however, is not described by the 
calculations}.
The measured values of $\gamma^{eff}_0$ agree, however, extremely well with the
corresponding values obtained from JETSET, as can be seen in fig.~4.
\section{Predictions on 2-particle angular correlations in jet cones}
Theoretical predictions \cite{ochs} concerning the emission of two 
partons with a relative
angle $\vartheta_{12}$ - within a cone with half opening angle $\Theta$ around
the jet axis - have been evaluated using two correlation functions defined 
as follows :
\begin{equation}
r(\vartheta_{12}) = \frac{\rho_2(\vartheta_{12})}{\rho_1\otimes
\rho_1(\vartheta_{12})}
\end{equation}
\begin{equation}
\tilde{r}(\vartheta_{12}) = \frac{\rho_2(\vartheta_{12})}{\bar{n}^2(\Theta)}
\end{equation}
with the correlation integrals \cite{lipa} $\rho_2(\vartheta_{12}) = 
\int_\Theta d^3k_1d^3k_2\rho_2(k_1,k_2)
\delta(\vartheta_{12} - \vartheta(k_1,k_2))$ and\\
$\rho_1\otimes\rho_1(\vartheta_{12}) = \int_\Theta d^3k_1d^3k_2\rho_1(k_1)
\rho_1(k_2)\delta(\vartheta_{12} - \vartheta(k_1,k_2))$  
where $\rho_1(k)$ is the single particle distribution 
and $\bar n(\Theta)$ is the mean multiplicity of partons emitted into the 
$\Theta$-cone. The quantities in eqns. (11) and (12) exhibit very different
structures. $\rho_2(\vartheta_{12})$ consists of 2 terms  
$\rho_2(\vartheta_{12}) = C_2(\vartheta_{12})$ +
$\rho_1\otimes\rho_1(\vartheta_{12})$ where only $C_2(\vartheta_{12})$ 
describes 
the genuine correlations, 
\mbox{$\rho_1\otimes\rho_1(\vartheta_{12})$}, on the other side, 
is obtained from
the single particle spectra. Consequently $r(\vartheta_{12})$ is given 
essentially by the normalized $C_2$-term whereas it turns out that 
the dominant term of
$\tilde{r}(\vartheta_{12})$ is given by   
$\rho_1\otimes\rho_1(\vartheta_{12})$.

Distinct predictions for $r(\vartheta_{12})$ and $\tilde{r}(\vartheta_{12})$ 
have been
evaluated which depend essentially only on the QCD parameters $\Lambda$ and 
$n_f$ \cite{ochs}. 

At high energy and for sufficiently large angles $\vartheta_{12} \leq 
\Theta$ the following power law is expected:
\begin{equation} r(\vartheta_{12}) =\left(\frac{\Theta}{\vartheta_{12}}\right)^{0.5
\gamma_{0}}  
\end{equation} 
and the scale determining $\gamma_0$ is again given by 
$Q \approx P\Theta$ (see section 3).

For asymptotically high energies the quantity
\begin{equation}
\frac{\ln(r(\vartheta_{12}))}{\sqrt{\ln\frac{P\Theta}{\Lambda}}} \approx
2\beta(\omega(\epsilon,2)-2\sqrt{1-\epsilon})
\end{equation}
with
\begin{equation}
\epsilon = \frac{\ln\frac{\Theta}{\vartheta_{12}}}{\ln\frac{P\Theta}{\Lambda}} 
{\rm{\quad , \qquad}} \beta^2 = 12(11-\frac{2}{3}n_f)^{-1}  
\end{equation}
and $\omega$ given by equ.~7,
is expected to be independent of the cone opening angle $\Theta$ and
primary momentum P, meaning that it is a scaling function.

The expected scaling properties of the quantity 
$\frac{\ln(r(\vartheta_{12}))}{\sqrt{\ln P\Theta /\Lambda}}$ (eqn.~(14)) 
are tested in figs.~5 and 6.

\begin{figure}
\vspace{-1.5cm}
\begin{minipage}[t]{5.5cm} 
\mbox{\epsfig{file=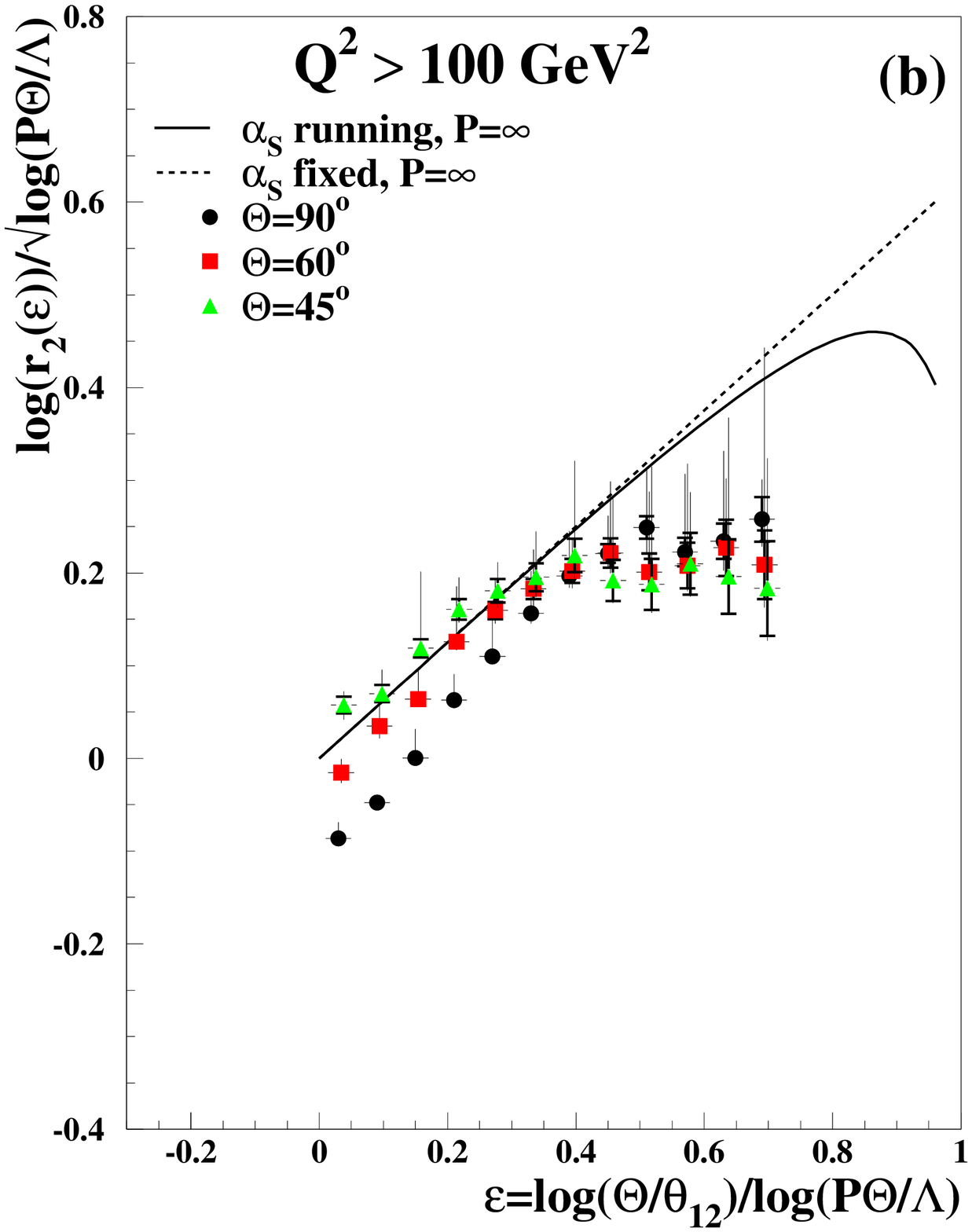,width=5.5cm}}
\end{minipage} \hfill 
\begin{minipage}[t]{5.5cm} 
\mbox{\epsfig{file=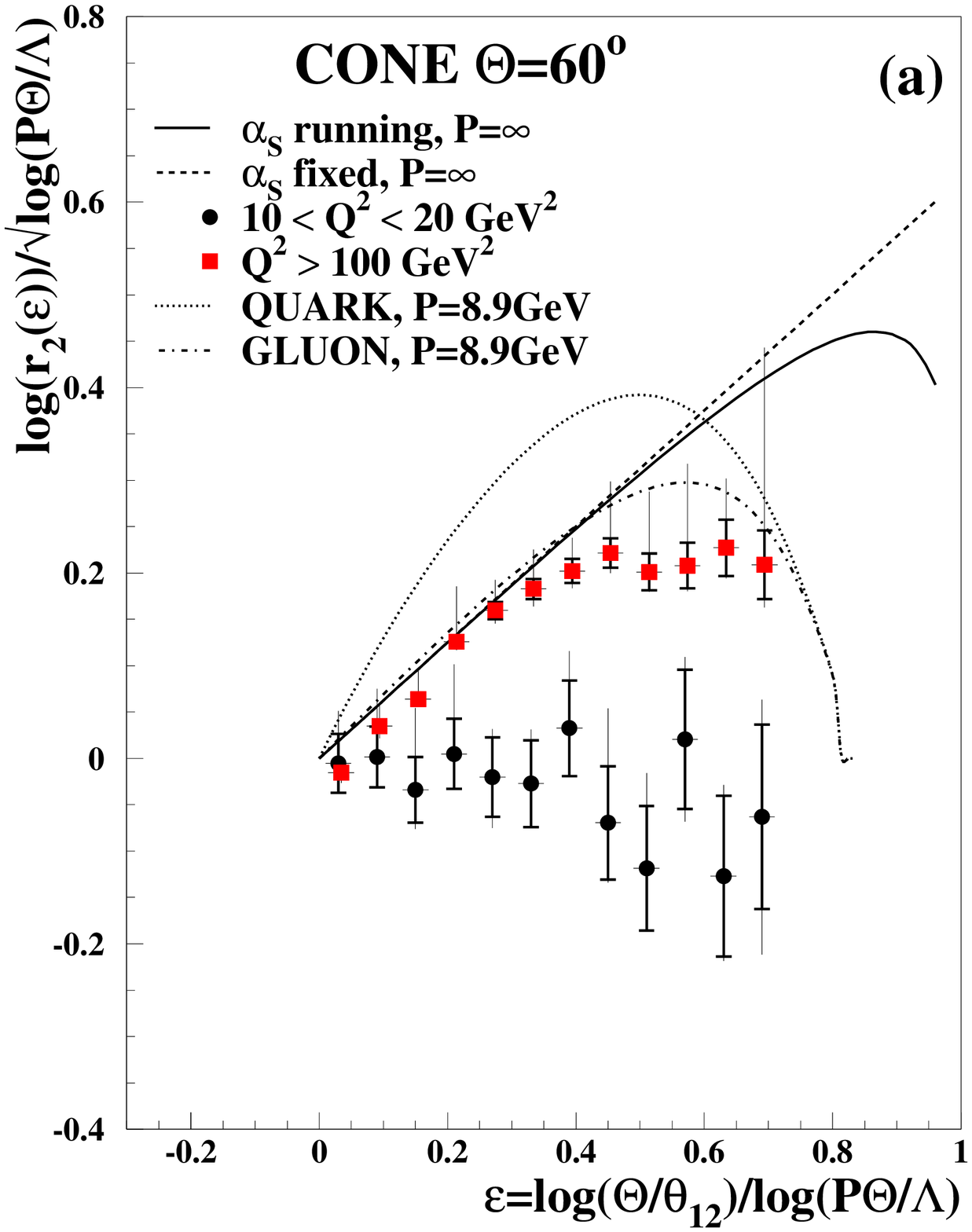,width=5.5cm}}
\end{minipage}
\caption[zeus2]{The $r_2$ correlation function (14) dependence on {\bf b)}
cone size $\Theta$ and {\bf a)} $Q^2$ in a lower energy range - and comparison
with QCD predictions \cite{ochs}.}
\end{figure}  

\begin{figure}
\vspace{-0.5cm}
\mbox{\epsfig{file=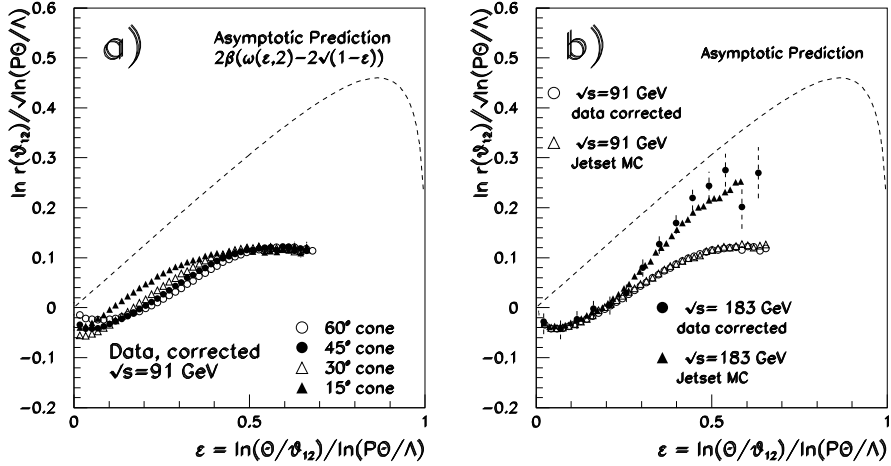,width=13cm}}
\vspace{-1.5cm}
\caption[]{
\footnotesize {\bf a)} The rescaled function $r_2$ (14) is 
plotted against the 
scaling variable
$\epsilon$ in order to test prediction \cite{ochs}, dashed line.  
{\bf b)} The corrected data  for $\Theta = 45^\circ$, at 91~GeV 
(open circles) and
at 183~GeV (full circles), using $\Lambda = 0.15~GeV$, 
are shown together with JETSET Monte Carlo calculations 
(open resp. full triangles).
}
\end{figure}
 
The dependence on the cone opening angle $\Theta$ is shown in figs.~5b (ZEUS)
and 6a (DELPHI) (left side of figs~5, 6). 
It can be seen that for $15^\circ \leq \Theta \leq 60^\circ$
the {\it dependence on $\Theta$ is very weak already at 
$\sqrt{s} = 91~GeV$}, 
in agreement with the predictions of eqn. (14). 
For small energies (fig~5b) there is no scaling for $\epsilon \leq 0.3$. 
The {\it shape predicted} by eqn.~(14)  
{\it differs appreciably from the measurement} at LEP
(fig.~6). 
The shape 
of the data is only similar to that predicted in the sense that it is 
rising and levelling off; the data are much smaller and flatter.  
There is a ``hook'' in data at small $\epsilon$ which is a reflection of
momentum conservation. It is almost absent, however, in the ZEUS data.
Thus at LEP energies the analytic QCD calculations do not
describe quantitatively the 2-particle angular 
correlations r($\vartheta_{12}$).
These differences are much less pronounced for the low energies
(solid lines in fig. 5), but this could 
be also due to the different way to define the jet axis.
The dependence of (14) on energy is shown in figs. 5a (ZEUS) and 6b 
(DELPHI) for a fixed
value of $\Theta$ (right hand sides of figs.~5,6).
The ZEUS data indicate an energy dependence only at lowest energies 
\cite{ze}.
This is in disagreement with the
the observations of DELPHI which observes {\it a strong energy 
dependence
also at the highest energies}.

In equs. (16), (17) a scaling function $Y(\epsilon)$ is defined, which is 
predicted to be 
independent of $\Theta$ and the primary momentum
P. 
\begin{equation} 
\tilde r(\epsilon)
=\vartheta_{12}\tilde r(\vartheta_{12})\ln\frac{P\Theta}{\Lambda} 
\end{equation} 
\begin{equation} Y(\epsilon) = -\frac{\ln(\tilde{r}(\epsilon)/b)}{2\sqrt{\ln(\frac{P\Theta}{\Lambda})}} =
2\beta(1-0.5\omega(\epsilon,2)) \quad,\quad b=2\beta\sqrt{\ln(\frac{P\Theta}{\Lambda})}
\end{equation} 

The expected scaling properties of $Y(\epsilon)$ are
tested in figs.~7 (ZEUS) and 8 (DELPHI). 
From $\sqrt{s} \approx 18~GeV$ upwards the {\it distributions 
agree very well with the prediction} in the region $\epsilon \geq 0.2$,
therefore exhibiting 
scaling both in angle $\Theta$ and energy. There is also good agreement 
with the coresponding JETSET simulations on both the partonic and
hadronic level,  
which {\it supports parton hadron
duality} - for the special function $Y(\epsilon)$. 
Note that no arbitrary
normalization has been applied in the above figures.

Similar to the case of factorial moments (fig.~4), values of $\gamma_{0}^{eff}$
have been extracted from $r(\vartheta_{12})$ by fitting the DELPHI data
to equ. (13) \cite{de2}. The corresponding values of $\alpha_s$ depending
on the cone opening angle $\Theta$ are shown in fig. 9a. Similar to the case
of factorial moments the data agree better with lower values of $\Lambda$.
This is at variance with the function $Y(\epsilon)$ where the data agree
best with the value $\Lambda \approx 0.3~GeV$ (fig.~9b). It has to be noted
that DLA takes only the leading singularities in all cases considered which
could lead to different redefinitions of the effective QCD parameters 
(e.g. $\Lambda_{eff}$) \cite{de2}.

\begin{figure}
\vspace{-1.0cm}
\begin{minipage}[t]{5.5cm} 
\mbox{\epsfig{file=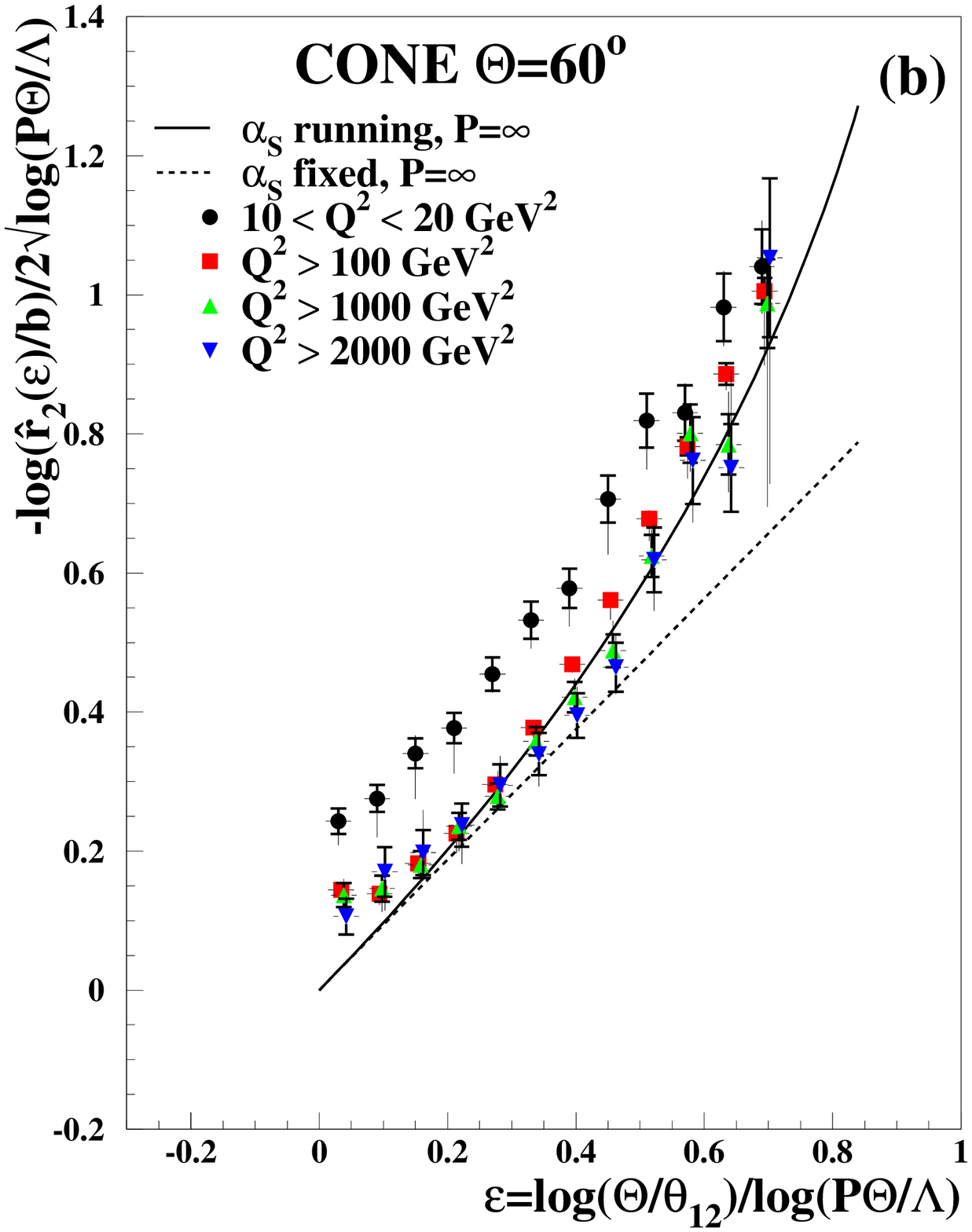,width=5.5cm}}
\end{minipage} \hfill 
\begin{minipage}[t]{5.5cm} 
\mbox{\epsfig{file=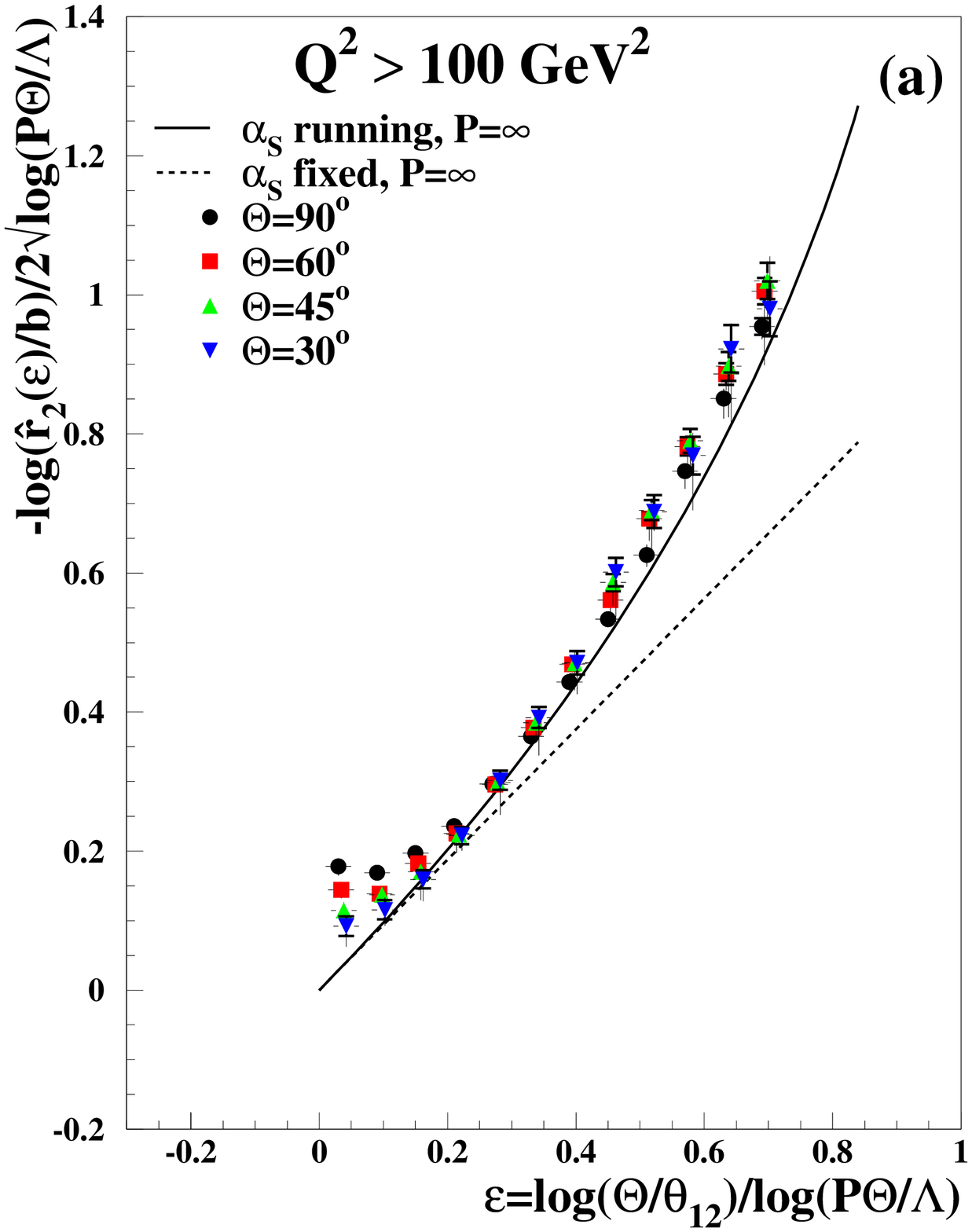,width=5.5cm}}
\end{minipage}
\caption[zeusrd]{The function Y($\epsilon$) (equ.~17) evolving with
{\bf b)} $Q^2$ and {\bf a)}
cone opening angle $\Theta$.}
\end{figure}  
\begin{figure}
\vspace{-0.5cm}
\mbox{\epsfig{file=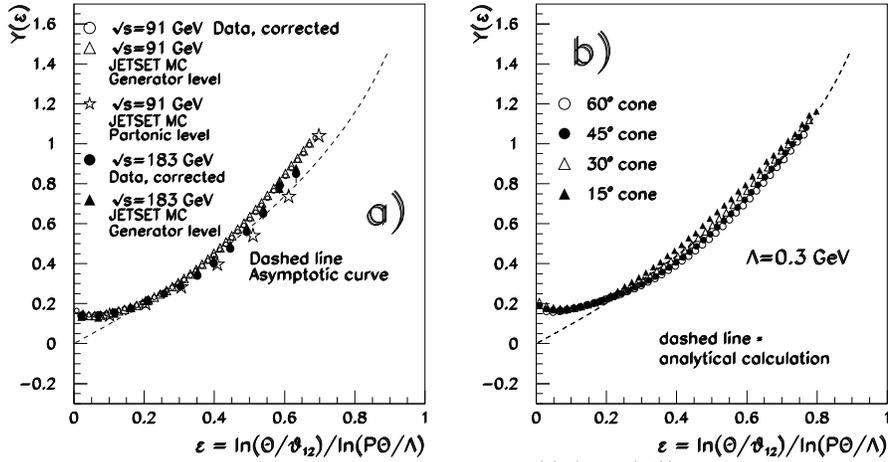,width=13cm}}
\vspace{-1.9cm}
\caption[]{
\footnotesize
An energy independent scaling function $Y(\epsilon)$ (eqn.~(17)) is 
extracted from the
2-body angular correlation function defined by eqn.~(12). The dashed lines
represent the asymptotic prediction eqn.~(17). Statistical and systematic
errors are smaller than 0.01 for the 91~GeV data. 
{\bf a)} The corrected data  for $\Theta = 45^\circ$, at 91~GeV (open circles) 
and at 183~GeV (full circles), 
using $\Lambda = 0.15~GeV$, are shown together with 
Monte Carlo calculations (open 
resp. full triangles). 
{\bf b)} Test of the $\Theta$-scaling behaviour of the data as predicted 
by eqn.~(17), using $\Lambda = 0.3~GeV$.
}
\end{figure}

\section{Summary and conclusions}
Present available QCD predictions on angular correlations, based on the 
DLA approximation, 
with first order relationship between $\alpha_s$ and $\Lambda$,
are checked,
comparing them to relevant results of three collaborations (DELPHI, L3 and
ZEUS). The experimental measurements, which agree generally well with
Monte Carlo simulations on hadronic level and
not so well with the partonic level
cover 2-particle angular correlations in cones 
and 1-dim and 2-dim multiplicity fluctuations in angular intervals. 
It turns out that functions which contain mainly single particle
terms ($Y(\epsilon)$) are predicted well in every respect. 
Also the extraction of QCD
parameters (i.e. $\Lambda_{eff}$) out of these measurements leads to reasonable
results. Differentially normalised correlations (r($\vartheta_{12})$,  
F($\frac{\Theta}{\vartheta}$) that contain 2 or 
more particles, however,
show often strong disagreements in shapes. 
The basic qualitative features, on the other hand, are
fulfilled. Here smaller values of $\Lambda_{eff}$ (resp. $\gamma_{0}^{eff}$ or 
$\alpha_{s}^{eff}$) are 
favoured as well
as $n_f$=3 rather than  $n_f$=5.
Up to now no set of QCD parameters will minimise simultaneously the various
discrepancies. 

\begin{figure}
\mbox{\epsfig{file=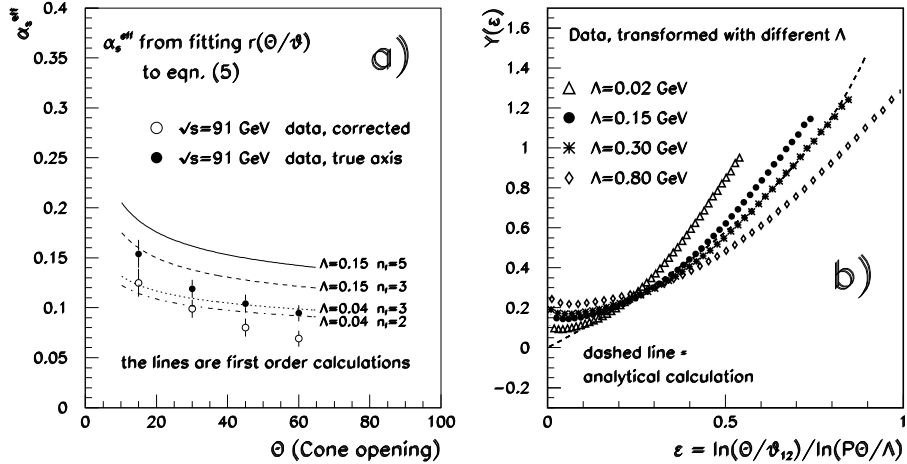,width=13cm}}
\vspace{-1.8cm}
\caption[]{
\footnotesize
{\bf a)} The measured values $\alpha_S^{\rm eff}$ ($\circ$) from eqn.~(13) for
different values of
$\Theta$ are compared with lowest order QCD relations equ.~(2),  
with $Q \approx P\Theta$, for different values of $\Lambda$ and $n_f$ . 
Applying Monte Carlo corrections for choosing the true axis (of the
initial parton) increases the value for $\alpha_S^{\rm
eff}$  ($\bullet$). The errors shown are systematic ones only, since the 
statistical errors are much
smaller. {\bf b)} Variation of the measured $Y(\epsilon)$ by 
choosing different values of $\Lambda$.
}
\end{figure}

The main shortcoming of the analytical calculations is the missing 
energy-momentum conservation adopted. Some "partial" introduction of
NLO corrections \cite{de1,l3} did not improve matters significantly.

Considering, however, that the analytical calculation does not use any
free parameter besides $\Lambda$ or $\alpha_s$, 
that they treat all kinds of correlations functions in an universal
manner,
that the asymptotic 
energies might
be still far away for some subtle functions and that,
nevertheless, the basic features of the predictions are always seen in the
experimental measurement, further theoretical efforts are encouraged.

We thank S.Chekanov, J.Fuster, W.Kittel and W.Ochs for valuable advice.


\section*{References}
\begin{thebibliography}{99}
\small{
\bibitem{parton} A. Basetto, M. Ciaffaloni, G. Marchesini, {\em Phys. Rep.} 
{\bf 100} (1983) 202;\\
Yu.L. Dokshitzer, V.A. Khoze, A.H. Mueller, S.I. Troyan, 
{\em Rev. Mod. Phys.} {\bf 60} (1988) 373 

\bibitem{lhd1} Ya.I. Azimov, Yu.L. Dokshitzer, V.A. Khoze, 
S.I. Troyan, {\em Z.Phys.}
{\bf C27} (1985) 65;
{\em Z. Phys.} {\bf C31} (1986) 231.

\bibitem{dla} V.S. Fadin, {\em Yad. Fiz.} {\bf 37} (1983) 408;\\ 
Yu.L. Dokshitzer, V.S.Fadin, V.A. Khoze, {\em Z. Phys.} {\bf C15} (1982) 325;\\ 
{\em ibid.} {\bf C18} (1983) 37.

\bibitem{ochs} 
W.Ochs, J.Wosiek {\em Phys.Lett.} {\bf B289} (1992) 159; \\
{\em Phys.Lett.} {\bf B304} (1993) 144;  
{\em Z. Phys.} {\bf C68} (1995) 269.


\bibitem{pesch} 
Ph.Brax, J.-L.Meunier and R.Peschanski, {\em Z.Phys.} {\bf C62} (1994) 649.


\bibitem{dremin} 
Yu.L.Dokshitzer and I.M.Dremin, {\em Nucl.Phys.} {\bf B402} (1993) 139.


\bibitem{x1} A.H. Mueller, {\em Phys. Rev.} {\bf D4} (1971) 150.

\bibitem{x2} K. Konishi, A. Ukawa and G. Veneziano, 
{\em Nucl. Phys.} {\bf B157} (1979) 45.

\bibitem{wien2} F.Mandl and B.Buschbeck, Proc. of the 24$^{\rm{th}}$ 
Internat. Symposium on
Multiparticle Dynamics, Vietri sul Mare, Italy, Sept. 1994, Eds. A. 
Giovannini, S. Lupia
and R. Ugoccioni, World Scientific 1995.

\bibitem{wien3} B.Buschbeck and F.Mandl, Proceedings of QCD 94, Montpellier,
France, July 1994, Ed. S.Narison, {\em Nucl. Phys. B (Proc. Suppl.)}
{\bf 39B,C} (1995), 150;\\
B.Buschbeck, P.Lipa, F.Mandl, Proc. of the 7$^{th}$ Internat. Workshop on
Multiparticle Production, Nijmegen, The Netherlands, 1996, Ed. R.Hwa et al.,
World Scientific;\\ 
B.Buschbeck, P.Lipa, F.Mandl, Proc. of the QCD 96, Montpellier, France,
July 1996, Ed. S. Narison, {\em Nucl. Phys. B (Proc. Suppl.)} {\bf 54A} 
(1997) 49.

\bibitem{de1}Multiplicity fluctuations in one- and two- dimensional angular
intervals compared with analytic QCD calculations, DELPHI collaboration,
to be published.

\bibitem{de2}Two-particle angular correlations in $e^+e^-$ interactions 
compared with QCD predictions, DELPHI collaboration, CERN-EP/98-138, 
accepted by Phys.Lett. B.

\bibitem{l3}
M.Acciarri et al (L3-Collab.), {\em Phys.Lett.} {\bf B428} 
(1998) 186. 

\bibitem{ze}Charged particle angular correlations in ep interactions at HERA,
ZEUS collaboration, paper 802 for ICHEP98, Vancouver, 23-29 July 1998 and
S. Chekanov, private communication.

\bibitem{delphi} P. Abreu et al. (DELPHI Collab.), {\em Phys. Lett.} {\bf B355} (1995) 415.

\bibitem{x} P.Abreu et al. (DELPHI Collab.), paper submitted to the ICHEP'98
conference, Vancouver, July 22-29, paper Nr.287. 

\bibitem{meun} J.--L. Meunier, private communication.

\bibitem{zeus} J.Breitweg et al., {\em Phys. Lett. B} {\bf 414} (1997) 428.

\bibitem{cumu}M.Kendall and A.Stuart, The Advanced Theory of Statistics,
Vol. 1, Charles Griffin, London 1977. 

\bibitem{lipa} P. Lipa et al., {\em Phys. Lett.} {\bf B285} (1992) 300;
P.Lipa er al., {\em Phys. Lett.} {\bf B301} (1993) 298.


}
\end{thebibliography}
\end{document}